\documentclass[a4paper,nobibnotes,nofootinbib]{revtex4}

\usepackage{amsmath,amssymb}
\usepackage{epsfig}

\newcommand{\g}{\gamma}
\newcommand{\f}{\frac}
\newcommand{\bea}{\begin{eqnarray}}

\newcommand{\intc}[1]{{\int\frac{d#1}{2i\pi}}}
\newcommand\lr[1]{{\left({#1}\right)}}

\begin{document}
\title{Saturation and forward jets at HERA}

\author{C. Marquet}\email{marquet@spht.saclay.cea.fr}
\author{R. Peschanski}\email{pesch@spht.saclay.cea.fr}
\affiliation{Service de physique th{\'e}orique, CEA/Saclay,
  91191 Gif-sur-Yvette cedex, France\footnote{%
URA 2306, unit{\'e} de recherche associ{\'e}e au CNRS.}}
\author{C. Royon}\email{royon@hep.saclay.cea.fr}
\affiliation{Service de physique des particules, CEA/Saclay,
  91191 Gif-sur-Yvette cedex, France}

\begin{abstract}

We analyse forward-jet production at HERA in the framework of the
Golec-Biernat and W\"usthoff saturation models. We obtain a good description
of the forward-jet cross-sections measured by the
H1 and ZEUS collaborations in the two-hard-scale region
($k_T\!\sim\!Q\!\gg\!\Lambda_{QCD}$) with  two different parametrizations
with either significant or weak saturation effects. The weak saturation
parametrization
gives a scale compatible with the one found for the proton structure
function $F_2$. We argue that Mueller-Navelet jets at
the Tevatron and the LHC could help distinguishing between both options.

\end{abstract}

\maketitle

\section{Introduction}
\label{1}

The saturation regime describes the high-density phase of
partons in perturbative QCD. It may occur for instance when the
Balitsky-Fadin-Kuraev-Lipatov (BFKL) QCD evolution equation \cite{bfkl} goes
beyond some energy  related to the unitarity limit
\cite{GLR,qiu,venugopalan,Balitsky:1995ub,levinbar,Mueller:2002zm}. On a
phenomenological ground, a well-known saturation model \cite{golec} by
Golec-Biernat and W\"usthoff (GBW) gives a parametrization  of
the proton structure functions already in the HERA energy range. It provides a
simple and elegant formulation of the transition to saturation. However, there
does not yet exist a clear confirmation of  saturation since  the same data can
well be explained within the conventional perturbative QCD framework
\cite{dglap}.

In fact, the study of the proton structure functions is a one-hard-scale
analysis since their
QCD properties are dominated by the evolution from a soft scale (the proton
scale)
to the hard scale of deep inelastic scattering. In order to favor the evolution
at fixed transverse scale, which is expected to lead more directly to
saturation, it seems
interesting to focus on two-hard-scale processes such as forward-jet production
at HERA, in the region where the transverse momentum of the jet $k_T$ is of the
order of the virtuality of the photon $Q^2$ and where the rapidity interval for
soft radiation is large. This kinematical configuration which was already
proposed \cite{dis} for testing the BFKL evolution is thus also a good testing
ground for saturation. The goal of our paper is then to formulate and study the
extension of the GBW model to forward jets at HERA.

Among the physics questions that we want to raise, we analyse whether
saturation effects can be sizeable in forward jets at HERA and whether
they are compatible or not with the GBW parametrization of $F_2.$ More
generally, we would like to compare potential saturation effects in one-scale
and two-scale processes. Indeed, in the two-scale process initiated by
$\g^*\g^*$ scattering, it has been suggested \cite{levin} that the saturation
scale could be different and in fact quite larger than the one of deep inelastic
scattering. However, another study of $\g^*\g^*$ scattering sticks to the same 
saturation scale as for $\g^*p$ scattering \cite{motyka}. We will discuss this 
point with forward-jet data which have better statistics and wider kinematical 
range than $\g^*\g^*$ scattering. If the saturation scale is universal as 
assumed in \cite{motyka}, then 
one should not expect to see stronger saturation effects in forward-jet data
than for $F_2;$ however if the saturation scale 
is higher for processes initiated by two hard scales as proposed in 
\cite{levin}, saturation effects could be more important. 

For HERA, the existence of two different saturation scales for one or two 
hard-scale processes would be an interesting feature.
This study is also of interest in the prospect of 
Mueller-Navelet jets \cite{navelet} at the Tevatron and the LHC where saturation 
with two hard scales could play a bigger role.

The plan of the paper is the following. In Section 2, we formulate the extension
of the GBW model for forward jets. In Section 3,
we present the
fitting method and give the results of the fits together with several tests of 
the
two saturation solutions that we obtain. Section 4 is devoted to
a discussion of our results and to predictions for Mueller-Navelet jets at
hadron
colliders that could discriminate between both solutions.

\begin{figure}[ht]
\begin{center}
\epsfig{file=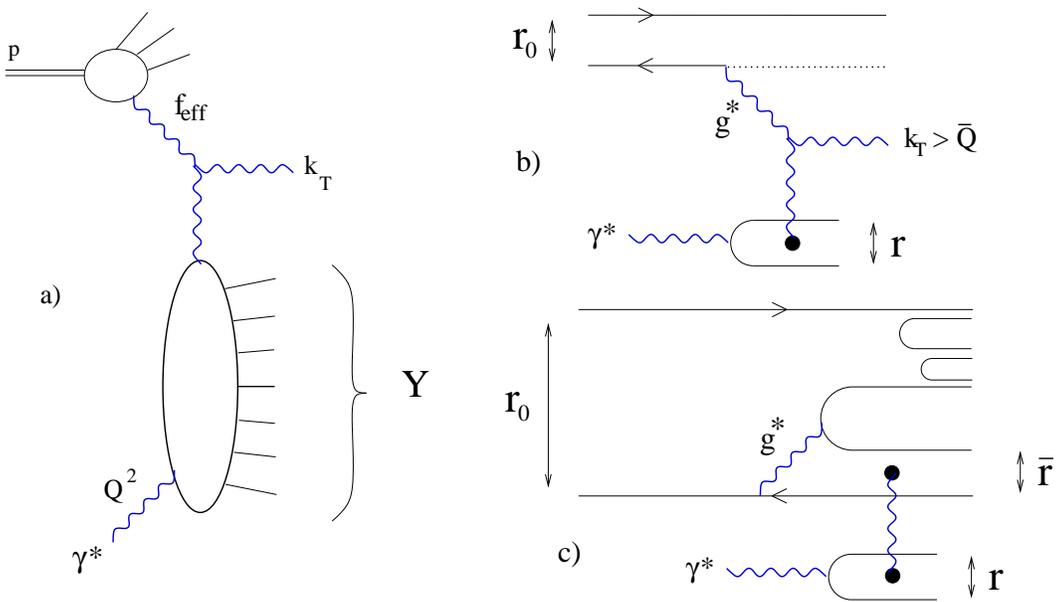,width=14cm}
\caption{{\it Forward-jet production in partonic and dipole representations.}
Fig 1a: Forward jets at HERA.
Fig 1b: Forward jet in $\gamma^*\!-\!onium$ scattering in the partonic
representation (at 1st order, {\it i.e.} with only one gluon exchanged). Fig 1c:
Forward jet
in $\gamma^*\!-\!onium$ scattering
in the dipole representation. $Y:$ rapidity
gap between the two hard probes. $k_T\!>\!\bar Q:$ jet tranverse momentum lower
bound. $Q:$
virtuality of the photon. The gluon-dipole couplings are sketched by
black points.}
\end{center}
\label{F1}
\end{figure}

\section{Formulation}
\label{2}

The original GBW model provides a simple formulation of
the dipole-proton cross-section in terms of a saturation scale
$R_0(Y)\!\sim\! e^{-\f{\lambda}2\left(Y-Y_0\right)}$ where
$Y\!=\!\log{1/x}$ is the total rapidity, $\lambda$ is the intercept and
$Y_0$ sets the absolute value. Varying $Q,$ $QR_0\!\gg\! 1$ corresponds to the
dilute limit
while when $QR_0\!<\!1$, the dipole-proton cross-section saturates to a finite 
limit
$\sigma_0$. In
two-hard-scale problems, one has now to deal  with dipole-dipole
cross-sections
which require an extension \cite{motyka} of the GBW parametrization. In the case
of
forward jets, one has
to combine the dipole-dipole cross-section with an appropriate definition of
the coupling of this cross-section to the forward jet. This has been proposed in
\cite{us}. We shall use this coupling in the analysis of forward jets at HERA.

The QCD cross-section for forward-jet production in a lepton-proton collision
reads
\begin{equation}
\f{d^{(4)}\sigma}{dxdQ^2dx_Jdk_T^2}=\f{\alpha}{\pi xQ^2}
\ f_{eff}(x_J,k_T^2)\left\{\lr{\f{d\sigma_T}{dk_T^2}+\f{d\sigma_L}{dk_T^2}}(1-y)
+\f{d\sigma_T}{dk_T^2}\f{y^2}2\right\},\label{one}\end{equation}
where $x$ and $y$ are the usual kinematic variables of deep inelastic 
scattering, $Q^2$ is the virtuality of the photon with
longitudinal (L) and  transverse (T) polarization, and $x_J$ is the jet 
longitudinal momentum fraction with respect to the proton.
$d\sigma_{T,L}/dk_T^2$ are 
the photon-parton hard differential cross-sections for the production of a
forward 
(gluon) jet
with transverse momentum $k_T\!\gg\!\Lambda_{QCD}$. The effective parton 
distribution function $f_{eff}$
has the following expression
\begin{equation}
f_{eff}(x_J,k_T^2)=g(x_J,k_T^2)+\f49\lr{q(x_J,k_T^2)+\bar{q}(x_J,k_T^2)}\ ,
\label{sf}\end{equation}
where $g$ (resp. $q$, $\bar{q}$) are the gluon (resp. quark, antiquark) 
structure
functions in the incident proton. $ k_T^2$ is chosen as the QCD factorization
scale.

Properly speaking, the forward-jet cross-section (1) is the leading
$\log{(1/x)}$ expression when one uses the BFKL formulation of
$d\sigma_{T,L}/dk_T^2.$
Let us show how one can extend this formalism when saturation corrections are
present, {\it e.g.} in the framework of the GBW model. The
differential hard cross-section reads
\begin{equation} \f{d\sigma_{T,L}}{dk_T^2}=
\left.-\f{\partial\sigma_{T,L}}{\partial\bar{Q}^2}(Q^2,\bar{Q}^2)
\right|_{\bar{Q}=k_T}
\end{equation}
where $\sigma_{T,L}$  are the cross-sections for the production of
a forward jet
with transverse momentum larger than $\bar{Q},$ expressed in the dipole 
framework
by \cite{us}
\begin{equation}
\sigma_{T,L}(Q^2,\bar{Q}^2)=\f{\pi^2\sigma_0}2\int d^2r\ d^2\bar{r}\
\phi^{\g}_{T,L}(r,Q^2) \
\phi^{J}(\bar{r},\bar{Q}^2)\left\{1-\exp{\lr{-\f{r_{eff}^2(r,\bar{r})}{4R_0^2(Y)
}}}\right\}\ .
\label{sigma}
\end{equation}
$\sigma_0$ times the term in brackets in formula (\ref{sigma}) is the GBW
dipole-dipole cross-section.
\begin{equation}
R^2_0(Y)=\f1{Q^2_0}\ e^{-\lambda\left(Y-Y_0\right)}\label{rad}
\end{equation}
is the saturation radius redefined  for the forward jet case, where
$Y\!=\!\log{x_J/x}$ is the rapidity interval available for
the forward-jet cross-section (see Fig 1), $\lambda$ is the saturation
intercept and  $Y_0$ is a parameter
defining the rapidity for which $R_0(Y_0)\!=\!1/Q_0\!\equiv\! 1$ GeV$^{-1}$. 
$\sigma_0$
is the value at which the dipole-dipole cross-section saturates. The effective 
radius 
$r_{eff}^2(r,\bar{r})$ is defined through the elementary dipole-dipole 
cross-section given by the two-gluon exchange \cite{mueller}
\begin{equation}
\sigma_{dd}(r,\bar{r})=\int d^2k f^0(k^2,r)f^0(k^2,\bar{r})=
2\pi\alpha^2_s r^2_{<}\lr{1+\ln\frac{r_{>}}{r_{<}}}\equiv
2\pi\alpha^2_s r^2_{eff}(r,\bar{r})\ ,\label{sdd}
\end{equation}
where $r^2_{<}$ (resp. $r^2_{>}$) is $\min{(r,\bar{r})}$ (resp. 
$\max{(r,\bar{r})}$),
$\alpha_s$ is QCD the coupling constant and $f^0(k^2,r)\equiv 
2\alpha_s(1-J_0(kr))/k^2$ is the dipole-gluon coupling. 
The functions $\phi^{\g}_{T,L}(r,Q^2)$ and $\phi^{J}(\bar{r},\bar{Q}^2)$ express
the couplings to the virtual photon and to the jet. $\phi^{\g}_{T,L}(r,Q^2)$ is
the known squared QED
wavefunction \cite{bjorken} of a photon in terms of $q\bar q$. 

In equation (\ref{sigma}), the definition of
$\phi^{J}(\bar{r},\bar{Q}^2)$ associated with
a forward jet with transverse momentum $k_T\!>\!\bar Q$ requires more care. The 
gluon jet is colored while the associated dipole $\bar r$ is not, see Fig.1c. 
$\phi^{J}$ is obtained by taking 
advantage of the
equivalence between the partonic and dipole formulations of forward-jet 
production when
the jet is emitted off an onium (see Fig 1b-1c, where the size of the incident 
onium 
is denoted $r_0$). Let us sketch the derivation of $\phi^J(\bar{r},\bar{Q}^2).$ 
Assuming the condition 
$1\,GeV^{-1}\!\gg\! r_0 \!\gg\! 1/\bar Q,$ the onium is small 
enough to 
allow for a perturbative QCD calculation but large enough with respect to the 
inverse transverse momentum of the forward jet.
At lowest pertubative order, the coupling of the system onium-jet to the gluon 
(interacting with the target photon) reads \cite{munier,robi,us}:
\begin{equation}
\alpha_s\log\frac 1{x_J} \int \frac{d^2\vec k_T}{\pi\vec k_T^2}\theta(\vec 
k_T^2-\bar{Q}^2)
f^0(|\vec k+\vec k_T|^2,r_0)\approx \left\{2\alpha_s\log\frac 1{x_J}
\log{\bar{Q}r_0}\right\}\int d^2\bar{r}\   
\frac{\bar{Q}}{2\pi\bar{r}}\ J_1(\bar{Q}\bar{r})\ f^0(k^2,\bar{r})\ . 
\label{phig}
\end{equation}
The function $f^0(k^2,\bar{r})$ factorizes as part of the dipole-dipole 
cross-section (\ref{sdd}).
The factor in brackets in (\ref{phig}) is the first order contribution of the  
Dokshitzer-Gribov-Lipatov-Altarelli-Parisi (DGLAP) 
gluon ladder \cite{dglap}, at  
the Double Leading Log (DLL) approximation. QCD factorization thus implies that 
it is 
included in the structure function of the incident particule. In the case of a 
proton it is absorbed in $f_{eff},$ see formula (\ref{sf}), by a 
redefinition of the factorization scale.
Therefore, the remaining term gives  
\begin{equation}
\phi^J(\bar{r},\bar{Q}^2) \equiv \frac{\bar{Q}}{2\pi\bar{r}}
\ J_1(\bar{Q}\bar{r})\ .
\label{bessel}
\end{equation}
Note that $\phi^{J}$ is an effective 
distribution that comes out from the calculation rather than a probability 
distribution as for $\phi^{\g}_{T,L}.$

Inserting in formula (\ref{sigma}) the known Mellin transforms
$\tilde{\phi}(\tau)\!=\!\int
d^2r \ (r^2Q^2)^{1-\tau}\phi(r,Q^2):$
\begin{equation}
\tilde{\phi}^J(\tau)=2^{2-2\tau}\frac{\Gamma(2-\tau)}{\Gamma(\tau)}\ ,
\hspace{1cm}
\tilde{\phi}^\g_{T,L}(\g)=\f{2N_c\alpha}{\pi}\sum_q e_q^2\f{1}{4^\g\g}
\f{\Gamma^2(1+\g)\Gamma^2(1-\g)\Gamma^2(2-\g)}
{\Gamma(2-2\g)\Gamma(2+2\g)(3-2\g)}
\lr{\begin{array}{cc}(1+\g)(2-\g)\\2\g(1-\g)\end{array}}
\label{gg}
\end{equation}
and after straightforward transformations we can express our results
in a double Mellin-transform representation:
\begin{eqnarray}
\f{d\sigma_{T,L}}{dk_T^2}=\f{\pi^2}{8Q^2k_T^2R_0^2(Y)}
\intc{\g}\ \tilde{\phi}^{\g}_{T,L}(\g)
(4Q^2R_0^2)^{\g}\intc{\tau}\ \tilde{\phi}^{J}(1\!-\!\tau)
(4k_T^2R_0^2)^{-\tau}\left\{\tau
\tilde{\sigma}(\tau,\g)\right\}\ ,\label{sigene}\\
0<Re(\tau),\;Re(\g),\;Re(\g-\tau)<1\ \nonumber\ ,
\end{eqnarray}
where
\begin{equation}
\tilde{\sigma}(\tau,\g)\equiv \sigma_0\int du^2\int d\bar{u}^2 u^{2\g-4}
\bar{u}^{-2\tau-2}\lr{1-e^{-r_{eff}^2(u,\bar{u})}}=\sigma_0
\f{2\Gamma{(\g-\tau)}}{1+\tau-\g}
\int_0^\infty du\ u^{-2\tau-1} \left[r_{eff}^2(1,u)\right]^{1+\tau-\g}\ .
\label{st}\end{equation}
Using (\ref{sdd}), one finally gets
\begin{equation}
\tilde{\sigma}(\tau,\g)=\sigma_0\ \f{2\Gamma{(\g-\tau)}}{1+\tau-\g}
\ \left\{\Psi(1,3+\tau-\g,2\tau)+\Psi(1,3+\tau-\g,2-2\g)\right\}\ \label{Psi},
\end{equation}
where the confluent hypergeometric function of Tricomi
$\Psi(1,a,b)$ can be expressed \cite{prud} in terms of incomplete Gamma
functions.

Note that, expanding the exponential function in the GBW dipole-dipole 
cross-section in
(\ref{sigma}), it is possible to single out order by order the contribution of
$r_{eff}^{2n}$ in formula (\ref{sigene}):
\begin{equation}
\tilde{\sigma}^{(n)}(\tau,\g)=2\sigma_0\f{(-1)^{n-1}}{n!}\ 2i\pi
\delta(\g-\tau+n-1)
\left\{\Psi(1,3+\tau-\g,2\tau)+\Psi(1,3+\tau-\g,2-2\g)\right\}\ .\label{Psin}
\end{equation}
There is an interesting physical interpretation of this expansion.
$\tilde{\sigma}^{(1)}$ corresponds to the dilute
limit of the GBW formula. Taking also into account $\tilde{\sigma}^{(2)}$
can be interpreted as saturation {\it a la} Gribov-Levin-Ryskin \cite{GLR}, {\it
i.e.} $\sim\! r_{eff}^2\!-\! r_{eff}^4/8R^2_0(Y)$. Note also that formulae
(\ref{sigene},\ref{st}) can also be used with the other
models for $r^2_{eff}$ proposed in the literature \cite{motyka}.

\section{Fitting forward jets}

For the fitting procedure, we will use a method allowing for
a direct comparison of the data with theoretical predictions. This method has
already been applied \cite{jets} for a BFKL parametrization of the same
$d\sigma/dx$ data at HERA. We shall extend this method to the GBW
parametrization (cf. Eq.(\ref{sigene})).

The published data depend on kinematical cuts (see \cite{h199,ze99}) which are
modeled by bin-per-bin {\it correction factors} that multiply the theoretical
cross-sections. The details of the method are as follows: {\bf i)}  for each
$x$-bin, one
determines \cite{cont} the average values of $Q^2$, $x_J$, $k_T$ from a
reliable Monte-Carlo simulation of the cross-sections using the {\it
Ariadne} program \cite{ariadne}; {\bf ii)}
one chooses a set of integration variables for $d^{(4)}\sigma$ (see (\ref{one}))
in such a way to match closely the experimental
cuts and minimize the variation of the cross-sections over the bin size, the
convenient choice of bins for forward jets
\cite{jets}  is
\begin{eqnarray}
\frac{d \sigma}{dx} = \int \left[Q^6 \frac{d^{(4)} \sigma}{dx dQ^2 dx_J d
k_T^2}\right] \times
\Delta \left(\frac {1} {Q^2} \right)\Delta x_J \Delta \left(\frac {k_T^2}
{Q^2}\right)\label{stringe}
\end{eqnarray}
(note the choice of the variable $k_T^2/Q^2$ which is well-suited for the
study of two-scale processes); {\bf iii)} one fixes the correction factors due
to the experimental cuts for each $x$-bin by a random simulation of the 
kinematic
constraints  with no dynamical input. The list of correction factors for
the H1 and ZEUS sets of data \cite{h199,ze99} and more details on the method are 
given in \cite{jets}.

Using these correction factors, we perform a
fit of formulae (\ref{one}) to the H1 and ZEUS data using the GBW
parametrization (\ref{sigene}-\ref{Psi}). The
free parameters are the saturation scale parameters
$\lambda$ and $Y_0$ (see (\ref{rad})) and the normalizations which we keep
independent for H1 and ZEUS. Note that they are related to the dimensionless
factors $\sigma_0 Q_0^2$. We also make the slight modification
$\sigma_0\!\rightarrow\!\alpha_S^2(k_T^2)\ \sigma_0$ with $\alpha_S$ running at 
one loop
with 4 active flavours and $\Lambda_{QCD} = 220$ MeV. Indeed, it is known
that leading-order BFKL fits are better when the coupling constant in the 
overall factor
is running. It turns out that with our GBW parametrization, the effect is much 
smaller
than in the BFKL case as will be discussed later on.
The obtained values of the parameters and the
$\chi^2$ of the fits are given in Table I and the resulting cross-sections
displayed\footnote{We follow the same procedure as in Ref.
\cite{jets}, namely
one H1 point at $k_T > 5$ GeV ($7.3$ 10$^{-4}$) and four ZEUS points
($x=4.$ 10$^{-4}$, and the three highest-x points) were not taken into account
in the fits. The three highest$-x$ points for ZEUS cannot be described by a
small$-x$ approach probably because the $x$-value is too high ($x> 10^{-2}$).
The other points cannot be described because of large correction
factors.
Note that similar discrepancies appear also in other types of fitting
procedures, {\it e.g.} in Ref. \cite{kwie}.} in Fig. 2. We obtain two different
solutions which can be characterised by different strength of saturation 
effects.
The $\chi^2$ per {\it dof} is quite good (we use only statistical errors to 
perform
the fit) in both cases.
The two solutions show similar
$\chi^2$ values and resulting cross-sections and we chose to plot only the
solution with higher saturation in Fig. 2 (1st line of table I, the other 
solution
would be indistinguishable on that same figure).
It displays the result of the fit (full line) together with
the contributions of the 1st ($\tilde\sigma^{(1)}$, dotted line) and 1st+2nd
($\tilde\sigma^{(1)}\!+\!\tilde\sigma^{(2)}$, dashed line) orders keeping the 
values
of the parameters found for the full solution. We can see that these orders
give distinguishable contributions.

\begin{table}
\begin{center}
\begin{tabular}{|c||c|c|c|c|c|c|c|} \hline
 fit & $\lambda$ & $Y_0$ &  ${\cal N}_{ZEUS}$ &  ${\cal N}_{H1}$
&
$\chi^2 (/dof)$ \\
\hline\hline
 sat. & 0.402 $\pm$ 0.036 $\pm$ 0.024
 & -0.82 $\pm$ 0.36 $\pm$ 0.01
 & 34.3 $\pm$ 8.5 $\pm$ 7.0
 & 31.7 $\pm$ 8.4 $\pm$ 8.7
 & 6.8 (/11)\\

  weak sat. & 0.370 $\pm$ 0.032 $\pm$ 0.015
  & 8.23 $\pm$ 0.48 $\pm$ 0.03
  & 1136  $\pm$ 272 $\pm$ 2
  & 1042 $\pm$ 238 $\pm$ 78& 8.3 (/11)\\
\hline
\end{tabular}
\end{center}\caption{{\it Results of the fits to the H1/ZEUS data for the GBW
model.}
We find two independent solutions showing either significant (1st line) or only 
weak (2nd line)
saturation parameters (see text). Statistical and systematic errors are 
indicated 
in that order for each parameter.}
\end{table}

\begin{figure}[ht]
\begin{center}
\epsfig{file=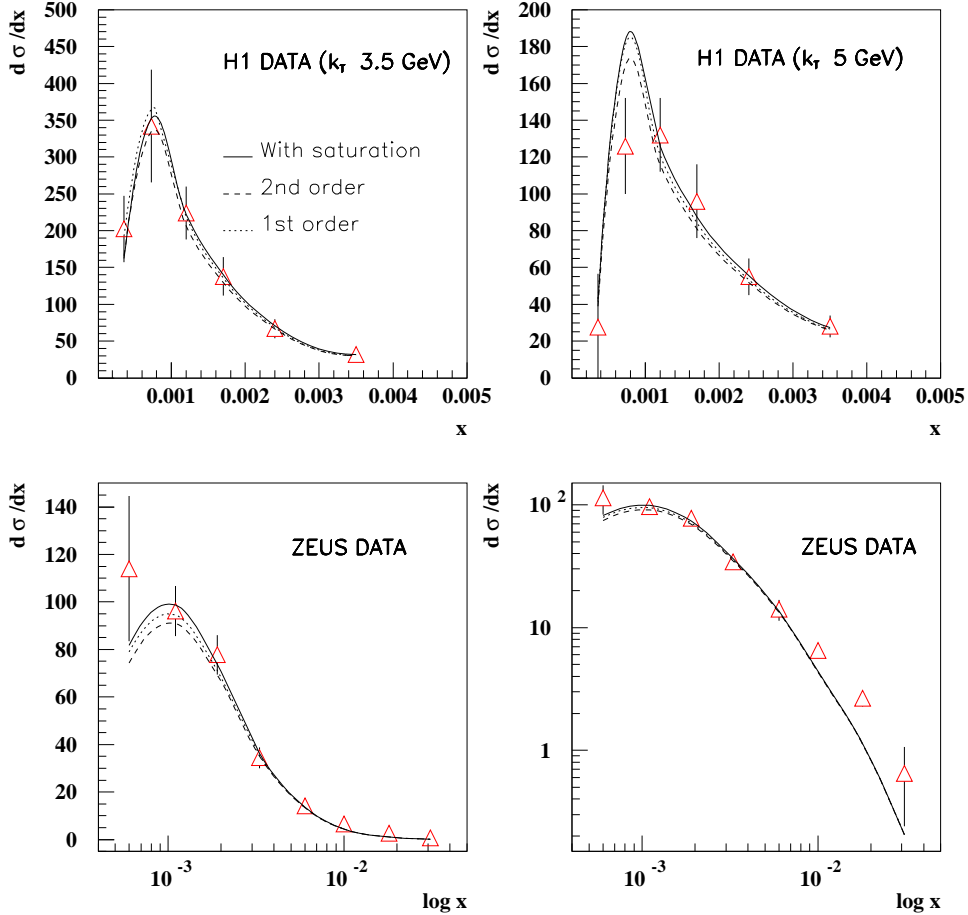,width=14cm,clip=true}
\caption{{\it Results of the fit with significant saturation.}
Upper left: H1 data $k_T\!>\! 3.5$ GeV, upper right: H1 data $k_T\!>\! 5$ GeV,
lower left and right:  ZEUS data in linear and logarithmic (showing
the discrepancy at high-$x$) scales. We display the result of the fit in full
lines, together with
the contribution of 1st and 1st+2nd orders (resp. dotted and dashed lines). The 
cross-sections are given in nanobarns.}
\end{center}
\label{F2}
\end{figure}

Let us now discuss each solution. The first solution (1st line of Table I)
shows a sizeably larger value of $\lambda$ than the $F_2$ parametrization
\cite{golec}
(cf. $\lambda^{F_2}\!=\! 0.288$), while the value of $Y_0$ is found to be 
completely
different (cf. $Y_{0}^{F_2}\!=\! 8.1$), leading to more significant saturation 
effects.
The second solution (2nd line of Table I) shows values
of $\lambda$ and $Y_0$ more compatible with the $F_2$ result, even if $\lambda$
remains somewhat larger. The normalizations for H1 and ZEUS data are found
to be compatible in both cases. By contrast, the second and absolute $\chi^2$ 
minimum
corresponds to fully
significant (if not very big) saturation effects. To analyse these features more 
in
detail, we study how the two $\chi^2$ minima appear order by order in the 
expansion
of the GBW cross-section using formula (\ref{Psin}). In Table II, we show the
results of the fits when we truncate the
expansion up to first, second and third order. The first order\footnote{In this 
case,
there is one less parameter as $e^{-Y_0}$ appears only in the normalization and 
is 
absorbed in ${\cal N}.$} (1st line in
Table II), which corresponds to the dilute limit (\ref{sdd}), is very
close to the 2nd minimum of the full model, showing that it indeed corresponds 
to weak
saturation effects. This also confirms that a mere BFKL
description \cite{jets} of $d\sigma/dx$, similar to the first order 
approximation (\ref{sdd}), fits well
the same data. At second order, which corresponds to the GLR
version of the model, only one minimum appears and it is closer to the
first minimum of Table I. At third order, one recovers the two minima observed 
with the full model.
It is clear that saturation corrections are needed to see the first minima 
appear and this confirms
that it involves significant saturation effects. The weak saturation solution 
should {\it a priori} be
present regardless of the number of terms in the expansion, yet it is not the 
case when the truncation
is done at second order. Our interpretation of this feature is that the 
cross-section (\ref{sigma}) is 
sensitive
to large dipoles and to the high$-r_{eff}$ behavior of the truncated 
dipole-dipole cross-section
even when the saturation radius $R_0$ is large; this is due to the fluctuations 
of the jet
dipole distribution (\ref{bessel}) around its mean value.

\begin{table}
\begin{center}
\begin{tabular}{|c||c|c|c|c|c|c|c|} \hline
 order & $\lambda$ & $Y_0$ & ${\cal N}_{ZEUS}$ &  ${\cal N}_{H1}$  & $\chi^2
(/dof)$ \\
\hline\hline
 1 & 0.372 $\pm$ 0.066 $\pm$ 0.015 & - & 53.7 $\pm$ 10.9
 $\pm$
7.6 &  49.3 $\pm$ 11.5 $\pm$ 10.4`& 8.3 (/12)\\
 2 sat & 0.452 $\pm$ 0.027 $\pm$ 0.027 & -1.71 $\pm$ 0.47 $\pm$ 0.01 & 20.2
$\pm$
5.8
$\pm$ 4.9 &
19.1 $\pm$ 5.2 $\pm$ 6.0
 & 6.8 (/11)\\
 3 (sat.) & 0.447 $\pm$ 0.025 $\pm$ 0.026 & -1.56 $\pm$ 0.33 $\pm$ 0.02 & 22.2
 $\pm$ 3.0
$\pm$ 5.3 & 20.9 $\pm$ 2.6 $\pm$ 6.6 & 6.8 (/12)\\
3 (weak) & 0.374 $\pm$ 0.028 $\pm$ 0.016 & 6.85 $\pm$ 0.32 $\pm$ 0.12 & 693
 $\pm$ 110
$\pm$ 78 & 637 $\pm$ 94 $\pm$ 35 & 8.2 (/12)\\
\hline
\end{tabular}
\end{center}\caption{{\it Results of the fits to the H1/ZEUS data for the GBW
models with different truncations.} 1st line: 1st order (no saturation),
2nd line: 2nd order ($\sim$ GLR model), 3rd and 4th line: 3rd order with either
significant  or  weak saturation effects.}
\end{table}

Let us make some additionnal comments. (i) We used $\alpha^2_S$ (in the overall 
factor) 
running at one loop, and for completion, we checked the results of the fits 
while
keeping $\alpha_S$ constant. We find the same quality of the fits but with the
effective saturation intercept enhanced by about 30\% for the first solution. 
(ii) We have
also studied the other
models for effective radii proposed in \cite{motyka} which lead to fits of the 
same
quality with comparable values for the paramaters. (iii) The physical
normalization (see equations (5) and (9)) is
${\cal N}\exp(- \lambda Y_0)$. We checked that we obtain a set of consistent
values around 50 for this physical normalisation for all fits. The disperse 
values of ${\cal N}$
are compensated by the values of $Y_0$.

In Fig 3. we plot the different saturation scales as a function of the physical
rapidity interval $Y.$ We also display the saturation scale parametrization
obtained by the GBW fit of the proton structure function $F_2$ \cite{golec}. The 
weak solution is compatible with the one from $F_2$ in the physical range from 
five to ten units in rapidity, the rapidity dependance is however somewhat 
stronger. This solution favors a universal saturation scale regardless of the 
presence of a soft scale in the problem, as proposed in \cite{motyka} where 
$\g^*\g^*$ data are described with the saturation scale obtained from $F_2.$ The 
strong saturation 
solution gives a curve that lies significantly above the weak and $F_2$ curves, 
indicating that in forward jets, the saturation scale is different and higher 
than for deep inelastic scattering, as claimed in \cite{levin}.

At this point it is important to compare the range of $Q_S^2\!\equiv\! 1/R_0^2$
with the experimental transverse momentum cuts. Notice that these
cuts (3.5 and 5 GeV) lie approximately in between the two possible saturation 
scales. It is thus clear why the saturation effects are different for both 
solutions.

\begin{figure}[ht]
\begin{center}
\epsfig{file=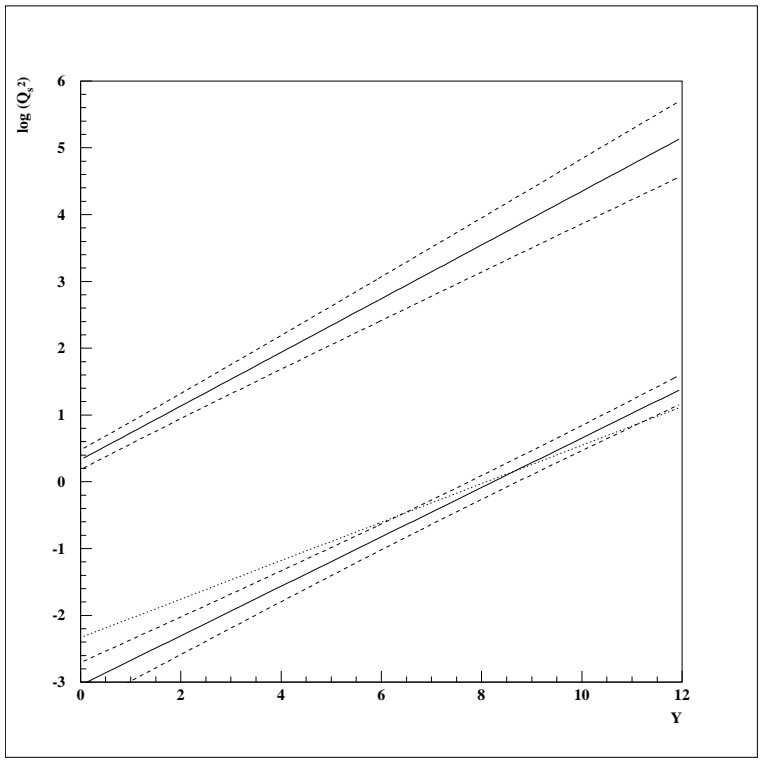,width=10cm,clip=true}
\caption{{\it Saturation scales $\log{Q_S^2/Q_0^2}\!=\!\lambda(Y-Y_0)$.}
We display the saturation scales for both solutions of our fits (full lines with 
errors 
indicated by dashed lines) and compare with the result for $F_2$ (dotted line). 
The weak 
saturation scale and the one for $F_2$ are compatible: lower curves. The other 
solution lies above, yielding stronger saturation effects.}
\end{center}
\label{F3}
\end{figure}

\section{Predictions for the Tevatron and the LHC}

The fact that the saturation corrections can be sizeable for two-scale processes
suggests that the predictions
for related hard cross-sections at hadronic colliders will show striking 
differences with the weak saturation case.
Let us for instance consider the Mueller-Navelet jet production \cite{navelet} 
at Tevatron and LHC following the approach of \cite{us}. Indeed, the
Mueller-Navelet jets are directly related to our discussion and, at the
level of the two-scale hard cross-section (\ref{sigma}), amounts to replacing
the photon probe by another forward jet. One writes

\begin{equation}
\sigma_{JJ}(k_{T1},k_{T2},\Delta\eta)=\f{N_c^2\sigma_0}{16}\int d^2r\
d^2\bar{r}\
\phi^{J}(r,k_{T1}^2) \
\phi^{J}(\bar{r},k_{T2}^2)\left\{1-\exp{\lr{-\f{r_{eff}^2(r,\bar{r})}
{4R_0^2(\Delta\eta)
}}}\right\}\ ,
\label{mnj}
\end{equation}
where $\Delta\eta$ is the rapidity interval between the two jets, and $k_{T1}$
and $k_{T2}$ are their lowest transverse momenta.
We thus consider cross-sections integrated over the transverse momenta of the 
jets
with lower bounds $k_{T1}$ and $k_{T2}$.
In order to appreciate more quantitatively the influence of saturation, it is
more convenient to consider the quantities ${\cal R}_{i/j}$ defined as
\begin{equation}
{\cal R}_{i/j} \equiv
\frac {\sigma(k_{T1},k_{T2},{\Delta \eta}_i)}{\sigma(k_{T1},k_{T2},{\Delta
\eta}_j)}\ ,
\label{Rij}
\end{equation}
 {\it i.e.} the cross-section ratios for two
different values of the rapidity interval. These ratios
display
in a clear way the saturation effects, they also correspond to possible
experimental  observables if one changes the center-of-mass energy
since they can be obtained from measurements at fixed values of
the jet light-cone momentum and thus are independent of the
parton densities the incident hadrons. Measuring two values of the cross-section 
for identical jet kinematics and different rapidity intervals $\Delta\eta_i$ and 
$\Delta\eta_j$ and dividing them gives the ratio ${\cal R}_{i/j}$. Such 
observables have actually been used for a study of Mueller-Navelet jets for 
testing BFKL predictions at the Tevatron \cite{goussiou,jets}. However, it is 
known that hadronization effects play a role in this kind of measurements and 
have to be taken into account \cite{schmidt}.

In Fig 4, we show the resulting ratios, when $k_{T1}\!=\! k_{T2}\!\equiv\! k_T,$ 
for ${\cal R}_{5/2}$ which corresponds to accessible rapidity intervals at
the Tevatron and ${\cal R}_{8/2}$ which corresponds to realistic rapidity 
intervals for the LHC. The curves are for both
saturation solutions together with the prediction obtained from the GBW
parametrization
of $F_2.$ At
high scales, both models lead to similar values of ${\cal R}$, larger than the
ones extracted \cite{us} from $F_2.$  This reflects the higher saturation
intercept systematically found with forward jets in both cases.

A clear difference between the two options (saturation being weak or not) appear 
in the drop due to saturation which occurs
for $k_T\!\sim\! Q_s.$ For both Tevatron and LHC kinematical configurations, the 
weak saturation solution would only give effects at rather and probably too low 
values of $k_T$ to be observable.
By contrast, the stronger saturation solution would give rise to saturation 
effects for $k_T$ of order $3\ GeV$ for the Tevatron and $30\ GeV$ for the 
LHC. While 
the effects could be marginal at Tevatron, the larger value of $k_T$ for the LHC 
provides a way to distinguish between the two solutions and to test the models. 
Hence the observation or not of saturation effects for Mueller-Navelet jets at 
LHC is a promising discriminant between the two options.

\begin{figure}[ht]
\begin{center}
\epsfig{file=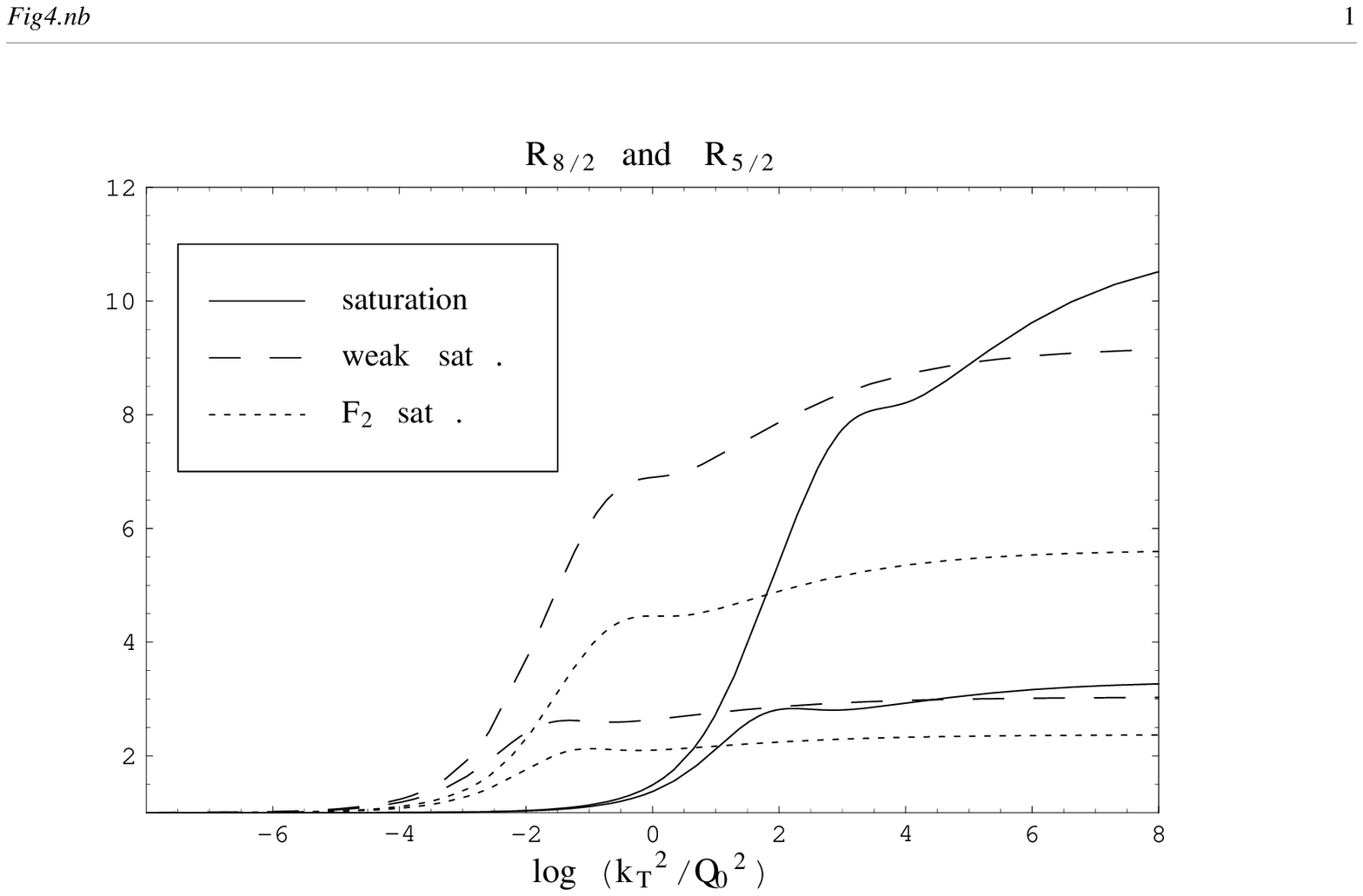,width=17cm,clip=true}
\caption{{\it Cross-section ratios ${\cal R}_{i/j}$.}
$R_{5/2}$ and $R_{8/2}$ are defined for rapidity intervals 2 and
5 for the Tevatron (3 lower curves),
and 2 and 8 for the LHC (3 upper curves). The full lines correspond to the 
significant saturation solution, first line of Table I. The dashed lines are for 
the weak saturation solution, second line of Table I. The dotted lines 
correspond to the saturation parametrization of $F_2$.}
\end{center}
\label{F4}
\end{figure}


\section{Conclusion}

Let us briefly summarize the main results of our study.
We described the published H1 and ZEUS forward-jet data using a saturation model
based on the GBW formalism. We find two possible fits to the data with either
significant or weak saturation. The weak saturation solution would be in favor 
of a universal QCD saturation scale. The solution with significant saturation 
effects would prove process dependent saturation scales.

Hence the confirmation or not of two different saturation scales, one for 
one-hard-scale processes 
({\it e.g. }for $F_2$), and one for two-hard-scale processes ({\it e.g. }for 
forward-jets) would suggest interesting theoretical questions, for instance 
whether the saturation scale is universaly related {\it e.g.} to $\Lambda_{QCD}$ 
or could be related to the hard probes initiating the process.

Both fits lead to different predictions at the
Tevatron, and even more different at the LHC where the effect is sizeable for 
larger
jet transverse momentum. The measurement of ${\cal R}_{i/j}$ would imply running
the accelerators at different center-of-mass energies. It would allow to test if
saturation is stronger for harder processes compared to deep inelastic 
scattering. It will also be interesting to test the two solutions using the 
recently presented preliminary H1 and ZEUS forward-jet and forward-pion data, 
when available.

\end{document}